\documentclass[twocolumn, floatfix, showpacs, prb]{revtex4}
\usepackage{amsmath}
\usepackage{graphicx}

\newcommand{\bra}[1]{\left<#1\right|}
\newcommand{\ket}[1]{\left|#1\right>}

\newcommand{\re}{\mathrm{e}}
\newcommand{\rd}{\mathrm{d}}

\begin{document}
\title{Counting statistics of coherent population trapping in quantum
  dots}
\author{C.W. Groth, B. Michaelis, and C.W.J. Beenakker}
\affiliation{Instituut-Lorentz, Universiteit
  Leiden, P.O. Box 9506, 2300 RA Leiden, The Netherlands}
\date{May 2006}
\begin{abstract}
  Destructive interference of single-electron tunneling between three
  quantum dots can trap an electron in a coherent superposition of
  charge on two of the dots.  Coupling to external charges causes
  decoherence of this superposition, and in the presence of a large
  bias voltage each decoherence event transfers a certain number of
  electrons through the device.  We calculate the counting statistics
  of the transferred charges, finding a crossover from sub-Poissonian
  to super-Poissonian statistics with increasing ratio of tunnel and
  decoherence rates.
\end{abstract}
\maketitle

\section{Introduction}
\label{intro}
The phenomenon of coherent population trapping, originating from
quantum optics, has recently been recognized as a useful and
interesting concept in the electronic context as
well.\cite{Bra04,Sie04} An all-electronic implementation, proposed in
Ref.\ \onlinecite{Mic06}, is based on destructive interference of
single-electron tunneling between three quantum dots (see Fig.\
\ref{fig:3dot}).  The trapped state is a coherent superposition of the
electronic charge in two of these quantum dots, so it is destabilized
as a result of decoherence by coupling to external charges.  In the
limit of weak decoherence one electron is transferred {\em on
  average\/} through the device for each decoherence event.

In an experimental breakthrough,\cite{Gus06,Gus06-2} Gustavsson et
al.\ have now reported real-time detection of single-electron
tunneling, obtaining the full statistics of the number of transferred
charges in a given time interval.  The first two moments of the
counting statistics give the mean current and the noise power, and
higher moments further specify the correlations between the tunneling
electrons.\cite{Bla00} This recent development provides a motivation
to investigate the counting statistics of coherent population
trapping, going beyond the first moment studied in Ref.\
\onlinecite{Mic06}.

Since the statistics of the decoherence events is Poissonian, one
might surmise that the charge counting statistics would be Poissonian
as well.  In contrast, we find that charges are transferred in bunches
instead of independently as in a Poisson process.  The Fano factor
(ratio of noise power and mean current) is three times the Poisson
value in the limit of weak decoherence.  We identify the physical
origin of this super-Poissonian noise in the alternation of two decay
processes (tunnel events and decoherence events) with very different
time scales---in accord with the general theory of Belzig.\cite{Bel05}
For comparable tunnel and decoherence rates the noise becomes
sub-Poissonian, while the Poisson distribution is approached for
strong decoherence.

The analysis of Ref.\ \onlinecite{Mic06} was based on the Lindblad
master equation for electron transport,\cite{Naz93,Gur98} which
determines only the average number of transferred charges.  The full
counting statistics can be obtained by an extension of the master
equation due to Bagrets and Nazarov\cite{Bag03} (without phase
coherence) and to Kie\ss{}lich et al.\cite{Kie06} (with phase
coherence).  In spite of the added complexity, we have found analytical
solutions for the second moment at any decoherence rate and for the
full distribution in the limit of weak or strong decoherence.

\section{Model}
\begin{figure}
\centering
\includegraphics[width=7cm]{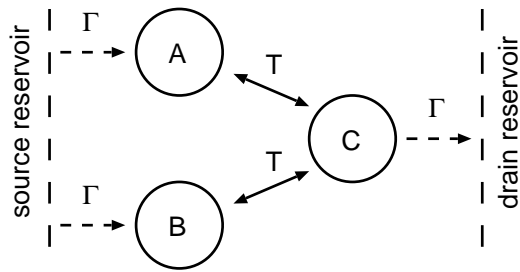}
\caption{Three quantum dots connected to a source and a drain
  reservoir.  Reversible transitions (rate $T$) and irreversible
  transitions (rate $\Gamma$) are indicated by
  arrows.}\label{fig:3dot}
\end{figure}
The system under consideration, studied in Ref.\ \onlinecite{Mic06},
is depicted schematically in Fig.\ \ref{fig:3dot}.  It consists of
three tunnel-coupled quantum dots connected to two electron
reservoirs.  In the limit of large bias voltage, which we consider
here, electron tunneling from the source reservoir into the dots and
from the dots into the drain reservoir is irreversible.  We assume
that a single level in each dot lies within range of the bias voltage.
We also assume that due to Coulomb blockade there can be at most one
electron in total in the three dots.  The basis states, therefore,
consist of the state $\ket{0}$ in which all dots are empty, and the
states $\ket{A}$, $\ket{B}$, and $\ket{C}$ in which one electron
occupies one of the dots.

The time evolution of the density matrix $\rho$ for the system is
given by the Lindblad-type master equation,\cite{Naz93,Gur98}
\begin{multline}
\label{eq:master}
\frac{\rd\rho}{\rd t} = -i[H,\rho]\\
+ \sum_{X=A,B,C,\phi_A,\phi_B,\phi_C} \hspace{-0.8cm}
\big(L_X\rho L^\dagger_X - \tfrac{1}{2}L^\dagger_X L_X\rho -
\tfrac{1}{2}\rho L^\dagger_X L_X \big).
\end{multline}
The Hamiltonian
\begin{equation}
\label{eq:H1}
H = T\ket{C}\bra{A} + T\ket{C}\bra{B} + \mathrm{H.c.}
\end{equation}
is responsible for reversible tunneling between the dots, with tunnel
rate $T$. For simplicity, we assume that the three energy levels in
dots $A$, $B$, and $C$ are degenerate and that the two tunnel rates
from $A$ to $C$ and from $B$ to $C$ are the same.  The quantum jump
operators
\begin{gather}
  \nonumber L_A = \sqrt{\Gamma}\ket{A}\bra{0},\quad
  \nonumber L_B = \sqrt{\Gamma}\ket{B}\bra{0}, \\
  L_C = \sqrt{\Gamma}\ket{0}\bra{C},
\end{gather}
model irreversible tunneling out of and into the reservoirs, with a
rate $\Gamma$ (which we again take the same for each dot).  Finally,
the quantum jump operators
\begin{equation}
L_{\phi_X} = \sqrt{\Gamma_\phi}\ket{X}\bra{X}, \quad X=A,B,C,
\end{equation}
model decoherence due to charge noise with a rate $\Gamma_\phi$.

As a basis for the the density matrix we use the four states
\begin{gather}
\nonumber \ket{e_0} = 2^{-1/2}(\ket{A}-\ket{B}),\\
\nonumber \ket{e_1} = 2^{-1/2}(\ket{A}+\ket{B}),\\
  \ket{e_2} = \ket{C},\quad \ket{e_3} = \ket{0}.
\end{gather}
If the initial state is $\ket{0}\bra{0}$ most of the coefficients of
$\rho$ remain zero.  We collect the five non-zero real variables in a
vector
\begin{equation}
v = (\rho_{00}, \rho_{11}, \rho_{22}, \rho_{33},
\mathrm{Im}\,\rho_{02})^\mathrm{T},
\end{equation}
whose time evolution can be expressed as
\begin{gather}
\label{eq:dotv}
\rd v/\rd t = Xv,\\
\addtolength{\arraycolsep}{-2pt}
X = \left(
\begin{array}{ccccc}
\Gamma_\phi/2 & -\Gamma_\phi/2 & 0 & \Gamma & 0 \\
-\Gamma_\phi/2 & \Gamma_\phi/2 & 0 & \Gamma & -2^{3/2}T \\
0 & 0 & -\Gamma & 0 & 2^{3/2}T \\
0 & 0 & \Gamma & -2\Gamma & 0 \\
2^{1/2}T & 0 & -2^{1/2}T & 0 & -\Gamma/2-\Gamma_\phi
\end{array}
\right).
\end{gather}

It is our goal to determine the full counting statistics, being the
probability distribution $P(n)$ of the number of transferred charges
in time $t$.  Irrelevant transients are removed by taking the limit
$t\to\infty$.  The associated cumulant generating function $F(\chi)$
is related to $P(n)$ by
\begin{equation}
\label{eq:cgf}
\exp[-F(\chi)] = \sum_{n=0}^\infty P(n) \exp(in\chi).
\end{equation}
From the cumulants
\begin{equation}
\label{eq:cumulants}
C_k=-(-i\partial_\chi)^kF(\chi)|_{\chi=0}
\end{equation}
we obtain the average current $I=eC_1/t$ and the zero-frequency noise
$S=2e^2C_2/t$, both in the limit $t\to\infty$.  The Fano factor is
defined as $\alpha = C_2/C_1$.

As described in Refs.\ \onlinecite{Bag03} and \onlinecite{Kie06}, in
order to calculate $F(\chi)$ one multiplies coefficients of the rate
matrix $X$ which are associated with tunneling into one of the
reservoirs (the right one in our case), by counting factors
$\re^{i\chi}$.  This leads to the $\chi$-dependent rate matrix
\begin{equation}
\addtolength{\arraycolsep}{-3pt}
\label{eq:lchi}
L(\chi) = \left(
\begin{array}{ccccc}
\Gamma_\phi/2 & -\Gamma_\phi/2 & 0 & \Gamma & 0 \\
-\Gamma_\phi/2 & \Gamma_\phi/2 & 0 & \Gamma & -2^{3/2}T \\
0 & 0 & -\Gamma & 0 & 2^{3/2}T \\
0 & 0 & \Gamma\re^{i\chi} & -2\Gamma & 0 \\
2^{1/2}T & 0 & -2^{1/2}T & 0 & -\Gamma/2-\Gamma_\phi
\end{array}
\right).
\end{equation}
The cumulant generating function for $t\to\infty$ can then be obtained
from the eigenvalue $\Lambda_\mathrm{min}(\chi)$ of $L(\chi)$ with the
smallest absolute real part,\cite{Bag03,Kie06}
\begin{equation}
\label{eq:cgf-eq-lambda}
F(\chi) = -t\Lambda_\mathrm{min}^{(\chi)}.
\end{equation}

\section{Results}
\subsection{Fano factor}
Low order cumulants can be calculated by perturbation theory in the
counting parameter $\chi$.  The calculation in outlined in App.\
\ref{sec:fano}.  For the average current we find
\begin{equation}
\label{eq:ssc}
I = \frac{4e\Gamma T^2}{\Gamma^2+14T^2
  + 2\Gamma\Gamma_\phi(1+2T^2/\Gamma_\phi^2)},
\end{equation}
in agreement with Ref.\ \onlinecite{Mic06}.  By calculating the noise
power and dividing by the mean current we obtain the Fano factor
\begin{eqnarray}
  \label{eq:fano}
  \nonumber  \alpha & = & \Big[\Gamma^4 + 148T^4
  + 4\Gamma^2(\Gamma_\phi^2 + 4T^2 + 12T^4/\Gamma_\phi^2) + \\
  & & (16T^2 + 2\Gamma^2)\beta \Big]
  \Big[\Gamma^2 + 14T^2 + \beta\Big]^{-2},\\
  \beta & = & 2\Gamma\Gamma_\phi(1 + 2T^2/\Gamma_\phi^2).
\end{eqnarray}

\begin{figure}
\includegraphics{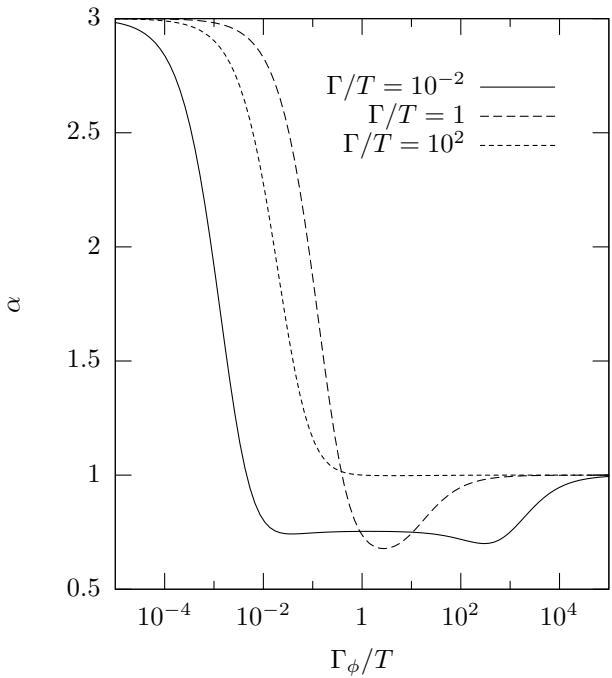}
\caption{Dependence of the Fano factor $\alpha$ on the normalized
decoherence rate $\Gamma_\phi/T$ for three values of $\Gamma/T$.}
\label{fig:fano}
\end{figure}

In Fig.\ \ref{fig:fano} the Fano factor has been plotted as a function
of $\Gamma_\Phi/T$ for three different values of $\Gamma/T$.  The
dependence of the Fano factor on the decoherence rate is nonmonotonic,
crossing over from super-Poissonian ($\alpha>1$) to Poissonian
($\alpha=1$) via a region of sub-Poissonian noise ($\alpha<1$).  To
obtain a better understanding of this behavior, we study separately
the regions of weak and strong decoherence.

\subsection{Weak decoherence}
For decoherence rate $\Gamma_\phi \ll \Gamma,T$ we have the limiting behavior
\begin{equation}
  \label{eq:weak}
  I \to e\Gamma_\phi,\quad
  \alpha \to 3 - \Gamma_\phi\left(\frac{17}{\Gamma}+\frac{\Gamma}{T^2}\right).
\end{equation}
Hence one charge is transferred on average per decoherence event, but
the Fano factor is three times the value for independent charge
transfers.

There exists a simple physical explanation for this behavior.  For
zero decoherence the system becomes trapped in the state $\ket{e_0}$.
The system is untrapped by ``decoherence events'', which occur
randomly at the rate $\Gamma_\phi$ according to Poisson statistics.
If $\Gamma_\phi$ is sufficiently small there is enough time for the
system to decay into the trapped state between two subsequent events,
so they can be viewed as independent.  The super-Poissonian statistics
appears because a single decoherence event can trigger the transfer of
more than a single charge.

The probability of $n$ electrons being transferred in total as a
consequence of one decoherence event is
\begin{equation}
R^1(n) = \frac{1}{2^{n+1}},
\end{equation}
since a decoherence event projects the trapped state $\ket{e_0}$ onto
itself or onto $\ket{e_1}$ with equal probabilities $1/2$ and each
electron subsequently entering the dots has a 50\% chance of getting
trapped in the state $\ket{e_0}$.

The number of electrons which have been transferred due to exactly $k$
decoherence events has distribution $R^k(n)$, the $(k-1)$th
convolution of $R^1(n)$ with itself.  We get
\begin{eqnarray}
  \nonumber
  R^k(n) & = &
  \frac{1}{2^{n+k}} \sum_{i_0=0}^n\sum_{i_1=0}^{i_0}\cdots\sum_{i_{k-2}=0}^{i_{k-3}} 1 \\
  & = & \frac{1}{2^{n+k}} \binom{n+k-1}{n}.
  \label{eq:rkn}
\end{eqnarray}
By definition,
\begin{equation}
R^0(n) = \delta_{n,0} = \left\{
\begin{array}{c}
1 \\
0
\end{array}
\,\mathrm{for}\,
\begin{array}{l}
n=0 \\
n>0
\end{array}
\right. ,
\end{equation}
being the distribution of the transferred charges after no decoherence
events have occurred.

The decoherence events in a time $t$ have a Poisson distribution,
\begin{equation}
P_{\mathrm{Poisson}}(k) = \re^{-t\Gamma_\phi}(t\Gamma_\phi)^k/k!.
\end{equation}
Combining with Eq.\ (\ref{eq:rkn}) we find the probability that $n$
electrons have been transferred during a time $t$,
\begin{eqnarray}
  P(n) & = & \sum_{k=0}^\infty P_{\mathrm{Poisson}}(k)R^k(n) \\
  \nonumber
  & = & \sum_{k=1}^\infty
  \frac{\re^{-t\Gamma_\phi}(t\Gamma_\phi)^k}{2^{n+k}k!}\binom{n+k-1}{n}
  + \re^{-t\Gamma_\phi}\delta_{n,0}.
\end{eqnarray}

The corresponding cumulant generating function is
\begin{equation}
\label{eq:cgf-low-decoh}
F(\chi)=t\Gamma_\phi-\frac{t\Gamma_\phi}{2-\re^{i\chi}},
\end{equation}
which gives rise to the cumulants
\begin{equation}
C_1=t\Gamma_\phi, \quad C_2=3t\Gamma_\phi, \quad C_3=13t\Gamma_\phi,
\end{equation}
in agreement with Eq.\ (\ref{eq:weak}).

The probability distribution (\ref{eq:cgf-low-decoh}) has been found
by Belzig in a different model.\cite{Bel05} As shown in that paper,
this superposition of Poisson distributions with Fano factor 3 arises
generically whenever there are two transport channels with very
different transport rates (in our case slow transport via the trapped
state $\ket{e_0}$, and fast transport via the untrapped state
$\ket{e_1}$).

\subsection{Strong decoherence}
We show that Poisson statistics of the transferred charges is obtained
for strong decoherence.  Consider the evolution equation
(\ref{eq:dotv}) of the system.  For $\Gamma_\phi \gg \Gamma,T$ the
coefficients $X_{00}, X_{01}, X_{10}$, and $X_{11}$ will ensure that
$v_0$ is equal to $v_1$ after a time which is short compared to the
other characteristic times of the system.  The trapped and the
non-trapped states will be equally populated.  Let us therefore define
\begin{equation}
v' = (\rho_{00}+\rho_{11}, \rho_{22}, \rho_{33},
\mathrm{Im}\,\rho_{02})^\mathrm{T}
\end{equation}
and use $\rho_{00}=\rho_{11}=v'_0/2$.
The evolution of $v'$ is governed by $\rd v'/\rd t=X'v'$, with
\begin{equation}
X' = \left(
\begin{array}{cccc}
0 & 0 & 2\Gamma & -2^{3/2}T \\
0 & -\Gamma & 0 & 2^{3/2}T \\
0 & \Gamma & -2\Gamma & 0 \\
2^{-1/2}T & -2^{1/2}T & 0 & -\Gamma/2-\Gamma_\phi
\end{array}
\right).
\end{equation}
The rate matrix $L'(\chi)$ is obtained by multiplying $X'_{12}$ by the
counting factor $\re^{i\chi}$.  An analytic expression can be found
for the smallest eigenvalue ${\Lambda'}_\mathrm{min}^{(\chi)}$ of
$L'(\chi)$, leading to the cumulant generating function
\begin{equation}
  F(\chi) = \frac{2T^2}{\Gamma_\phi}t\left(1-\re^{i\chi}\right)
\end{equation}
of a Poisson distribution.

\section{Conclusion}
In conclusion, we have shown that coherent population trapping in a
purely electronic system has a highly nontrivial statistics of
transferred charges.  Depending on the ratios of decoherence rate and
tunnel rates, both super-Poissonian and sub-Poissonian statistics are
possible.  We have obtained exact analytical solutions for the
crossover from sub- to super-Poissonian charge transfer, and have
calculated the full distribution in the limits of weak and strong
decoherence.  We hope that the rich behavior of this simple device
will motivate experimental work along the lines of Ref.\
\onlinecite{Gus06} and \onlinecite{Gus06-2}.

\acknowledgments
We thank L. Ament for useful discussions.  This research was supported
by the Dutch Science Foundation NWO/FOM. CWG acknowledges a
Cusanuswerk Fellowship.

\appendix
\section{Derivation of the Fano factor}
\label{sec:fano}
To derive the result (\ref{eq:fano}) for the Fano factor it is
sufficient to know the cumulant generating function to second order in
$\chi$.  The eigenvalues of the rate matrix $L(\chi)$ defined in Eq.\
(\ref{eq:lchi}) have the expansion
\begin{equation}
\label{eq:lambda}
\lambda = \lambda_0 + \lambda_1 \chi + \lambda_2 \chi^2 + O(\chi^3).
\end{equation}
We seek the eigenvalue with the smallest real part in absolute value.
That eigenvalue has $\lambda_0=0$.  We also express the eigenvector
$w$ corresponding to $\lambda$ and the matrix itself in a power series
in $\chi$:
\begin{eqnarray}
w & = & w_0 + w_1 \chi + w_2 \chi^2 + O(\chi^3), \\
L & = & L_0 + L_1 \chi + L_2 \chi^2 + O(\chi^3).
\end{eqnarray}
Inserting the above expansions into the eigenvalue equation $Lw =
\lambda w$ yields the following relationships of respectively zeroth,
first and second order:
\begin{eqnarray}
L_0w_0 & = & 0, \\
L_1w_0 + L_0w_1 & = & \lambda_1w_0, \\
L_2w_0 + L_1w_1 + L_0w_2 & = & \lambda_2w_0 + \lambda_1w_1.
\end{eqnarray}
The coefficients $L_k$ are known, while $w_k$ and $\lambda_k$ remain
to be found by solving these equations sequentially.  The first two
cumulants then follow from
\begin{equation}
C_1 = -it\lambda_1,\quad C_2 = -2t\lambda_2.
\end{equation}
In an analogue way it is possible to calculate higher cumulants.

\bibliographystyle{apsrev.bst}

\end{document}